\documentclass[twocolumn,showpacs,floatfix,prl]{revtex4}%
\usepackage{graphicx}%
\usepackage{amsmath}%
\usepackage{amsfonts}%
\usepackage{amssymb}
\usepackage{bm}

\def\s{{\sigma}}
\def\e{{\epsilon}}
\def\k{{ {\bm k} }}
\def\p{{ {\bm p} }}
\def\q{{ {\bm q} }}

\def\0{{ {\bm 0} }}

\def\a{{\alpha}}

\allowdisplaybreaks[4]

\begin{document}
\title{
Sign-reversing orbital polarization in the nematic phase of FeSe \\
due to the $C_2$ symmetry-breaking in the self-energy 
}
\author{
Seiichiro \textsc{Onari}$^{1,2}$, 
Youichi \textsc{Yamakawa}$^{3}$, and 
Hiroshi \textsc{Kontani}$^{3}$
}

\date{\today }

\begin{abstract}
To understand the nematicity in Fe-based superconductors,
nontrivial $\k$-dependence of the 
orbital polarization ($\Delta E_{xz}(\k)$, $\Delta E_{yz}(\k)$)
in the nematic phase,
such as the sign reversal of the orbital splitting
between $\Gamma$- and X,Y-points in FeSe, 
provides significant information.
To solve this problem,
we study the spontaneous symmetry breaking with respect to
the orbital polarization and spin susceptibility 
self-consistently.
In FeSe, due to the sign-reversing orbital order,
the hole- and electron-pockets are elongated
along the $k_y$- and $k_x$-axes respectively, consistently with experiments.
In addition, an electron-pocket splits into two Dirac cone Fermi pockets
with increasing the orbital polarization.
The orbital-order in Fe-based superconductors originates from 
the strong positive feedback 
between the nematic orbital order and spin susceptibility.

\end{abstract}

\address{
$^1$ Department of Physics, Okayama University,
Okayama 700-8530, Japan
\\
$^2$ Research Institute for Interdisciplinary Science, Okayama University
\\
$^3$ Department of Physics, Nagoya University,
Furo-cho, Nagoya 464-8602, Japan
}
 

\sloppy

\maketitle


The spontaneous symmetry breaking from $C_4$- to $C_2$-symmetry,
so called the electronic nematic transition,
is one of the fundamental unsolved electronic properties 
in Fe-based superconductors.
To explain this nematicity, both the spin-nematic scenario
\cite{Fernandes,DHLee,Chubukov,QSi,Valenti,Fanfarillo}
 and the orbital order scenario
\cite{Kruger,PP,WKu,Onari-SCVC,Onari-SCVCS,Text-SCVC,JP}
have been studied intensively.
Above the structural transition temperatures $T_{\rm str}$,
large enhancement of the electronic nematic susceptibility
predicted by both scenarios
\cite{Fernandes,Onari-SCVC}
is actually observed by the measurements of the 
softening of the shear modulus $C_{66}$
\cite{Fernandes,Yoshizawa,Bohmer,Onari-SCVC},
Raman spectroscopy, 
\cite{Gallais,Kontani-Raman,Khodas-Raman},
and in-plane resistivity anisotropy $\Delta\rho$
\cite{Fisher}.

To investigate the origin of the nematicity,
FeSe ($T_{\rm c}=9$ K) is a favorable system since the 
electronic nematic state without magnetization is realized 
below $T_{\rm str}=90$ K down to 0 K.
Above $T_{\rm str}$,
the antiferromagnetic fluctuations is weak and $T$-independent
according to the NMR 
\cite{Ishida-NMR,Dresden-NMR}
and neutron scattering
\cite{FeSe-neutron1,FeSe-neutron2,FeSe-neutron3}
studies,
in contrast to the sizable spin fluctuations 
above $T_{\rm str}$ in LaFeAsO 
 \cite{NMR-LaFeAsO}
and BaFe$_2$As$_2$
 \cite{NMR-BaFe2As2}.
This fact means that the magnetic instability is not a necessary 
condition for the electronic nematic state.
In contrast to the smallness of the spin fluctuations,
large nematic susceptibility is measured by $C_{66}$ and $\Delta\rho$ in FeSe.
Based on the orbital-spin fluctuation theory,
called the self-consistent vertex-correction (SC-VC) theory,
the development of the strong orbital fluctuations in FeSe are explained
even when the spin-fluctuations are very small,
consistently with experimental reports in FeSe
 \cite{FeSe-Yamakawa}:
The strong orbital fluctuations originate from
the Aslamazov-Larkin vertex correction (AL-VC)
that describes the orbital-spin mode-coupling
\cite{Onari-SCVC}.
The nematic CDW in cuprate superconductor
also originates from the AL-VC 
\cite{Yamakawa-CDW,Tsuchiizu-CDW}.

The nontrivial electronic state below $T_{\rm str}$
gives a crucial test for the theories proposed so far.
In the orthorhombic phase with $(a-b)/(a+b) \sim 0.3$\%,
large orbital-splitting $|E_{xz}-E_{yz}|$ of order $50$ meV 
is observed  at X,Y-points by ARPES studies in BaFe$_2$As$_2$
\cite{ARPES-Shen},
NaFeAs
\cite{ARPES-NaFeAs},
and FeSe
\cite{FeSe-ARPES1,FeSe-ARPES2,FeSe-ARPES22,FeSe-ARPES3,FeSe-ARPES4,FeSe-ARPES5,FeSe-ARPES6,FeSe-ARPES7,FeSe-ARPES8}.
Especially, noticeable deformation of the Fermi surfaces (FSs)
with $C_2$-symmetry is realized in FeSe, because of the 
smallness of the Fermi momenta.
In FeSe, Ref. \cite{FeSe-ARPES6} reports that
the orbital splitting $E_{xz}-E_{yz}$ is positive at $\Gamma$-point,
whereas it is negative at X,Y-points.
This sign-reversing orbital splitting 
is not realized in the non-magnetic orthorhombic phase in NaFeAs
 \cite{ARPES-NaFeAs}.
In addition, the e-FS1 at X-point is
deformed to two Dirac cone Fermi pockets in thin-film FeSe
 \cite{FeSe-ARPES5,FeSe-ARPES8}.
The aim of this study is to explain these nontrivial
electronic states in the orbital-ordered states
based on the realistic multiorbital Hubbard model.

Microscopically, the orbital order
is expressed by the symmetry breaking in the self-energy.
In the mean-field level approximations, however,
the self-energy is constant in $\k$-space unless 
large inter-site Coulomb interactions are introduced \cite{JP}.
For this reason, we have to study the non-local correlation effect
beyond the mean-field theory,
based on the realistic Hubbard model with on-site Coulomb interaction.
We will show that the strong positive feedback between 
the nematic orbital order
and $C_2$-symmetric spin susceptibility plays the essential role.

In this paper, we study the origin of the orbital order
in Fe-based superconductors,
by considering the cooperative symmetry breaking 
between the self-energy and  spin susceptibility 
self-consistently.
Experimentally observed strong $\k$-dependent 
orbital polarization ($\Delta E_{xz}(\k)$, $\Delta E_{yz}(\k)$)
is given by the non-local self-energy with $C_2$-symmetry.
In the FeSe model,
we obtain the sign-reversing orbital-splitting
$E_{xz}-E_{yz}$ between $\Gamma$-point and X,Y-points reported in Refs. 
\cite{FeSe-ARPES6}.
In addition, two Dirac-cone Fermi pockets emerge around X-point
when the Coulomb interaction is larger \cite{FeSe-ARPES5,FeSe-ARPES8}.
Thus, important key experimental electronic properties
in the orbital-ordered phase are satisfactorily explained.

Hereafter, we denote the five $d$-orbital
$d_{3z^2-r^2}$, $d_{xz}$, $d_{yz}$, $d_{xy}$, $d_{x^2-y^2}$ as $l=1,2,3,4,5$.
We study the realistic eight-orbital $d$-$p$ Hubbard models
\begin{eqnarray}
H_{\rm M}(r)=H_{\rm M}^0+rH_{\rm M}^U \ \ \
\mbox{(M = LaFeAsO and FeSe)},
\label{eqn:Ham}
\end{eqnarray}
where 
$H_{\rm M}^U$ is the first-principles screened Coulomb potential
for $d$-orbitals in Ref. \cite{Arita}:
The averaged Coulomb interaction 
${\bar U}\equiv \frac{1}{5}\sum_{l=1}^5U_{l}=7.21$ eV for FeSe
is much larger than ${\bar U}=4.23$ eV for LaFeAsO,
since the number of the screening bands,
by which ${\bar U}$ is reduced, is small in FeSe  \cite{Arita}.
In contrast, the averaged Hund's coupling 
$\bar{J}\equiv \frac{1}{10}\sum_{l>m}J_{l,m}$
is similar in all compounds
since the screening on the Hund's coupling is small.
For this reason, the ratio 
$\bar{J}/\bar{U}=0.0945$ in FeSe is much smaller than the ratio 
$\bar{J}/\bar{U}=0.134$ in LaFeAsO.
The factor $r(<1)$ is introduced to adjust the spin fluctuation strength.

\begin{figure}[!htb]
\includegraphics[width=.9\linewidth]{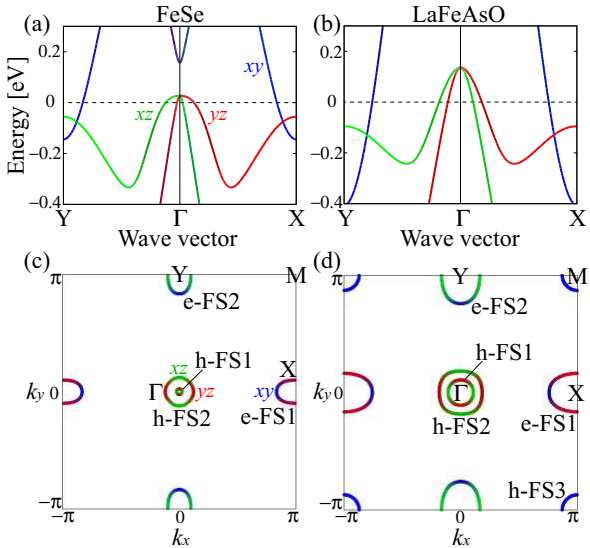}
\caption{
(color online)
Bandstructures of the eight-orbital models for
(a) FeSe and (b) LaFeAsO in the unfolded Brillouin zone 
at $T=50$ meV.
FSs for the FeSe and LaFeAsO models are shown in (c) and (d), respectively.
The colors correspond to 2 (green), 3 (red), and 4 (blue), respectively.
}
\label{fig:FS}
\end{figure}

$H_{\rm M}^0$ in Eq. (\ref{eqn:Ham}) is the first-principle tight-binding model 
introduced in Ref. \cite{FeSe-Yamakawa}.
The band dispersion $E_\k^\a$ is the solution of 
${\rm det}({\hat z}^{-1}\e+\mu-{\hat h}_M^0(\k))=0$,
where ${\hat h}^0_{\rm M}(\k)$ is the kinetic term
and ${\hat z}^{-1}$ is the diagonal mass-enhancement factor:
$({\hat z}^{-1})_{l,l'}\equiv 1/z_l\cdot\delta_{l,l'}$.
For LaFeAsO, we put ${\hat z}^{-1}={\hat 1}$.
For FeSe, we put $1/z_4 =1.6$ and $1/z_l=1$ ($l\ne4$)
to represent the strong renormalization of the $d_{xy}$-orbital band
\cite{FeSe-ARPES1}. 
Figures \ref{fig:FS} show the bandstructures  
and the FSs in the FeSe and LaFeAsO models
with $n_{\rm tot}=2\sum_{\a,\k}f(E_\k^\a)=12$.
In FeSe, each Fermi pocket is very shallow
\cite{BEC}.
The detail explanation for $H_{\rm M}^0$
is given in the Supplemental Material (SM) A \cite{SM}.

In this paper,
we calculate the $\k$-dependence of the self-energy 
using the self-consistent one-loop approximation,
which have been applied to various single-orbital models
\cite{FLEX}
and multi-orbital models
\cite{Text-SCVC} in literature.
The Green function in the orbital basis is 
\begin{eqnarray}
{\hat G}(k)=({\hat z}^{-1}i\e_n+\mu-{\hat h}^0_{\rm M}(\k)
-\Delta{\hat \Sigma}(\k))^{-1},
\label{eqn:Gr}
\end{eqnarray}
where $k=(\k,\e_n=(2n+1)\pi T)$ and
$\Delta{\hat \Sigma}(\k)$ is symmetry breaking self-energy.
and ${\hat z}^{-1}$ is the diagonal mass-enhancement factor.
The self-energy in the one-loop approximation 
\cite{Text-SCVC,FLEX}
is given as
\begin{eqnarray}
\Sigma_{l,l'}(k)=\Sigma_{l,l'}^{\rm H}
+T\sum_{q,m,m'} V_{l,m;l',m'}(q)G_{m,m'}(k-q),
\label{eqn:Self}
\end{eqnarray}
where $\Sigma_{l,l'}^{\rm H}=-\sum_{m,m'}\Gamma_{l,l';m',m}^c\Delta n_{m,m'}$
is the Hartree term:
$\Delta n_{m,m'}\equiv \langle c_{i,m,\s}^\dagger c_{i,m',\s}\rangle
-\langle c_{i,m,\s}^\dagger c_{i,m',\s}\rangle_0$.
The non-local interaction ${\hat V}(q)$ in Eq. (\ref{eqn:Self}) is given as
$\displaystyle 
\frac32 {\hat \Gamma}^s{\hat \chi^s}(q){\hat \Gamma}^s
+\frac12 {\hat \Gamma}^c{\hat \chi^c}(q){\hat \Gamma}^c
-\frac12 \bigl[{\hat \Gamma}^c{\hat\chi}^0(q){\hat \Gamma}^c
+{\hat \Gamma}^s{\hat\chi}^0(q){\hat \Gamma}^s
-\frac14 ({\hat \Gamma}^s+{\hat \Gamma}^c){\hat\chi}^0(q)
({\hat \Gamma}^s+{\hat \Gamma}^c) \bigr]
$,
where 
${\hat \chi}^{s,c}(q)={\hat\chi}^0(q)(1-{\hat \Gamma}^{s,c}{\hat \chi^0(q)})^{-1}$
and $\chi_{l,l',m,m'}^0(q)= -T\sum_k G_{l,m}(k+q)G_{m',l'}(k)$.
Here, ${\hat \Gamma}^{c(s)}$ is the bare Coulomb interaction 
for the charge (spin) channel given in the SM A \cite{SM}.

From Eq. (\ref{eqn:Self}), 
the $B_{1g}$-type symmetry breaking self-energy 
included in Eq. (\ref{eqn:Gr}), which is orbital-diagonal, is derived as
%
\begin{eqnarray}
\Delta \Sigma_{l}(\k)= 
{\rm Re} \left\{\Sigma_{l,l}(\k,\e_n)-\Sigma_{l}^{\rm A_{1g}}(\k,\e_n)
\right\}_{i\e_n\rightarrow 0}
\label{eqn:Self2}
\end{eqnarray}
for $l=2,3$, where
$\Sigma_{l}^{\rm A_{1g}}(k)\equiv(\Sigma_{l,l}(k)+\Sigma_{5-l,5-l}(k'))/2$ 
($\k'=(k_y,k_x)$) is the $\rm A_{1g}$-component of the self-energy.
In the present study, 
we calculate Eqs. (\ref{eqn:Self})-(\ref{eqn:Self2}) self-consistently
\cite{comment}.
We will show that $\Delta \Sigma_{l}$ emerges
due to the strong positive feedback between 
the nematic orbital order
and $C_2$-symmetric spin susceptibility:
That is, near the nematic transition,
infinitesimally small nematic orbital order enhances the 
spin susceptibility at $\q=(\pi,0)$, and the enhanced 
spin-fluctuation-mediated interaction ${\hat V}(q)$
in turn enlarges the orbital order.

\begin{figure}[t]
\includegraphics[width=.99\linewidth]{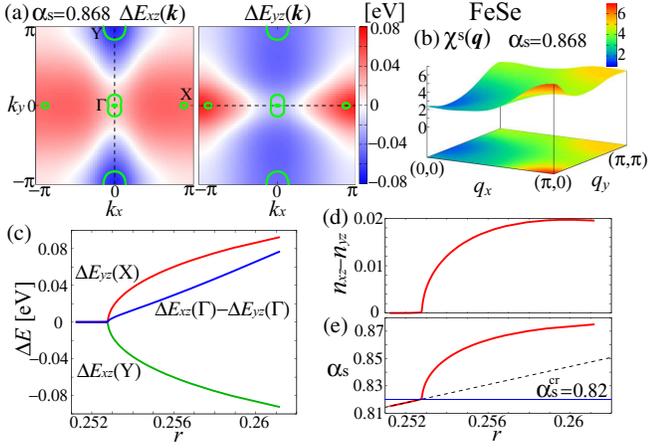}
\caption{
(color online)
(a) Orbital sign-reversing polarization
($\Delta E_{xz}(\k)$, $\Delta E_{yz}(\k)$) and 
(b) $\chi^s(\q)$ in the FeSe model
for $\a_S=0.868$ ($r=0.257$).
The realized FSs are shown by the green lines.
The splitting $E_{xz}-E_{yz}$ 
at $\Gamma$-point is positive, as observed in Ref. \cite{FeSe-ARPES6}.
(c) The polarization at $\Gamma$-, X- and Y-points.
(d) $\Delta n \equiv n_{xz}-n_{yz}$ and (e) $\a_S$ for $0.261>r>0.251$.
}
\label{fig:FeSe}
\end{figure}

First, we study the FeSe model,
using $64\times64$ $\k$-meshes and $512$ Matsubara frequencies
at $T=50$ meV.
In FeSe, the orbital order with
$\Delta E_{xz}(\k) \equiv \Delta \Sigma_2(\k)$ and
$\Delta E_{yz}(\k) \equiv \Delta \Sigma_3(\k)$ emerges
when the spin Stoner factor $\a_S$,
which is the maximum eigenvalue of $\hat{\Gamma}^s\hat{\chi}^0(\q)$,
is larger than $0.82$ ($r>0.253$).
The magnetic order is realized when $\a_S=1$.
In Figs. \ref{fig:FeSe} (a) and (b), we show respectively
the obtained orbital polarization
($\Delta E_{xz}(\k)$, $\Delta E_{yz}(\k)$) and the spin susceptibility 
$\chi^s(\q)\equiv \sum_{l,m}\chi^s_{l,l;m,m}(\q)$ for $\a_S=0.868$.
Due to the positive $\Delta E_{yz}({\rm X})$,
the e-FS1 around X-point is modified to
the pair of the Dirac-cone Fermi pockets.
Another electron-pocket around Y-point, e-FS2, is enlarged
by the negative $\Delta E_{xz}({\rm Y})$.
We stress that $\Delta E_{xz(yz)}(\k)$ changes its sign 
along the $k_{y(x)}$-axis shown by the broken line in Fig. \ref{fig:FeSe} (a).
Due to this sign reversal,
the outer hole-pocket (h-FS2) is elongated along the $k_y$-axis,
as observed experimentally \cite{FeSe-ARPES6}.
Due to the orbital polarization, 
$\a_S$ increases and  $\chi^s(\q)$ shows the $C_2$ anisotropy
$\chi^s(\pi,0) > \chi^s(0,\pi)$ shown in Fig. \ref{fig:FeSe} (b)
\cite{Kontani-softening,Fanfarillo}.

In Fig. \ref{fig:FeSe} (c),
we show the $r$-dependences of the orbital polarization at 
$\Gamma$-, X- and Y-points.
With increasing $r$, the spin-fluctuation-driven
orbital order appears as a second-order transition at $\a_S^{\rm cr}=0.82$.
The relations $\Delta E_{xz}({\rm Y})<0$ and $\Delta E_{yz}({\rm X})>0$ hold
in the ordered state.
In FeSe, the orbital splitting $E_{xz}-E_{yz}$ 
at $\Gamma$-point is positive,
so h-FS2 is elongated along the $k_y$-axis.
Such sign-reversing orbital order does not occur in the LaFeAsO model
(see Fig. \ref{fig:LaFeAsO}).
Since $d_{xz}$- and $d_{yz}$-orbitals are exchanged by $\pi/2$-rotation,
the obtained order
($\Delta E_{xz}(\k), \Delta E_{yz}(\k)$) belongs to the $B_{1g}$ representation (=$d$-wave),
in spite of $\Delta E_{xz}(\k)\neq \Delta E_{yz}(\k)$
\cite{FeSe-ARPES4,JP}.

Figure \ref{fig:FeSe} (d) shows the
difference $\Delta n \equiv n_{xz}-n_{yz}$:
$\Delta n \sim +10^{-2}$ will induce the small lattice deformation 
$(a-b)/(a+b)\approx0.2$\% in FeSe due to small $e$-ph interaction.
When  $\Delta n \ne0$,
the spin Stoner factor $\a_S$ is strongly enlarged
as shown in Fig. \ref{fig:FeSe} (e),
consistently with the enhancement of spin fluctuations 
observed by NMR \cite{Ishida-NMR,Dresden-NMR}
and neutron \cite{FeSe-neutron1,FeSe-neutron2} measurements.
For a fixed $r$, $\Delta n$ shows a mean-field-type second-order 
$T$-dependence below $T_{\rm str}$ 
\cite{FeSe-Yamakawa}.

\begin{figure}[!htb]
\includegraphics[width=.9\linewidth]{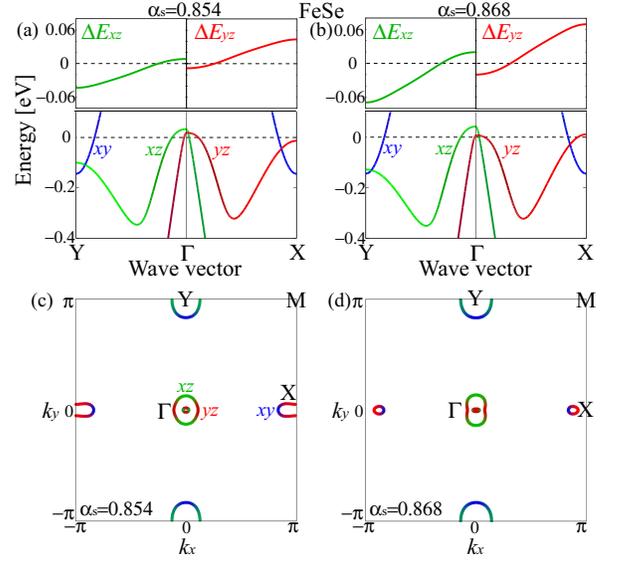}
\caption{
(color online)
Orbital polarizations and
bandstructures in the FeSe model along 
${\rm Y}\rightarrow \Gamma \rightarrow{\rm X}$
for (a) $\a_S=0.854$ and (b) $\a_S=0.868$.
The corresponding FSs at $T=50$ meV are shown in (c) and (d), respectively.
}
\label{fig:SCVC}
\end{figure}

In Fig. \ref{fig:SCVC},
we display the $C_2$ bandstructures obtained in the FeSe model for 
(a) $\a_S=0.854$ and (b) $\a_S=0.868$.
In both cases,
$\Delta E_{yz}(k_x,0)$ and $\Delta E_{xz}(0,k_y)$ show sign reversal, 
and $E_{xz}-E_{yz}>0$ at $\Gamma$-point.
The shape of the FSs in Fig. \ref{fig:SCVC} (c) for $\a_S=0.854$
(Fig. \ref{fig:SCVC} (d) for $\a_S=0.868$)
is consistent with the FSs observed in bulk FeSe (thin-film FeSe).
This result indicates that the electron correlation 
in thin-film FeSe is slightly stronger,
consistently with the higher $T_{\rm str}\approx120$K in thin-film FeSe.

In real FeSe, 
the inner pocket sinks under the Fermi level due to the spin-orbit interaction.
Similar single hole-FS model is introduced in the SM B \cite{SM} 
by shifting the $d_{xy}$-orbital level at $\Gamma$-point.
It is verified that the 
sign-reversing orbital polarization emerges in the FeSe model
with single hole-FS in the SM B \cite{SM}.

\begin{figure}[!htb]
\includegraphics[width=.99\linewidth]{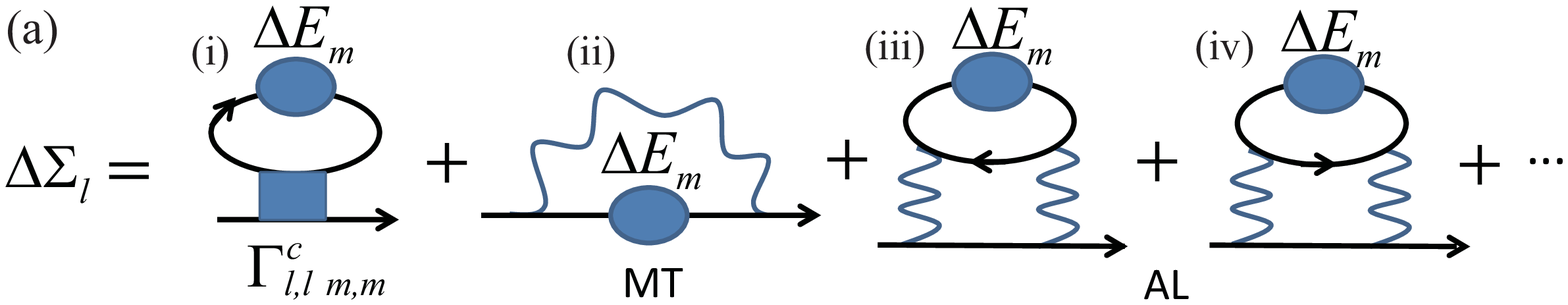}
\includegraphics[width=.9\linewidth]{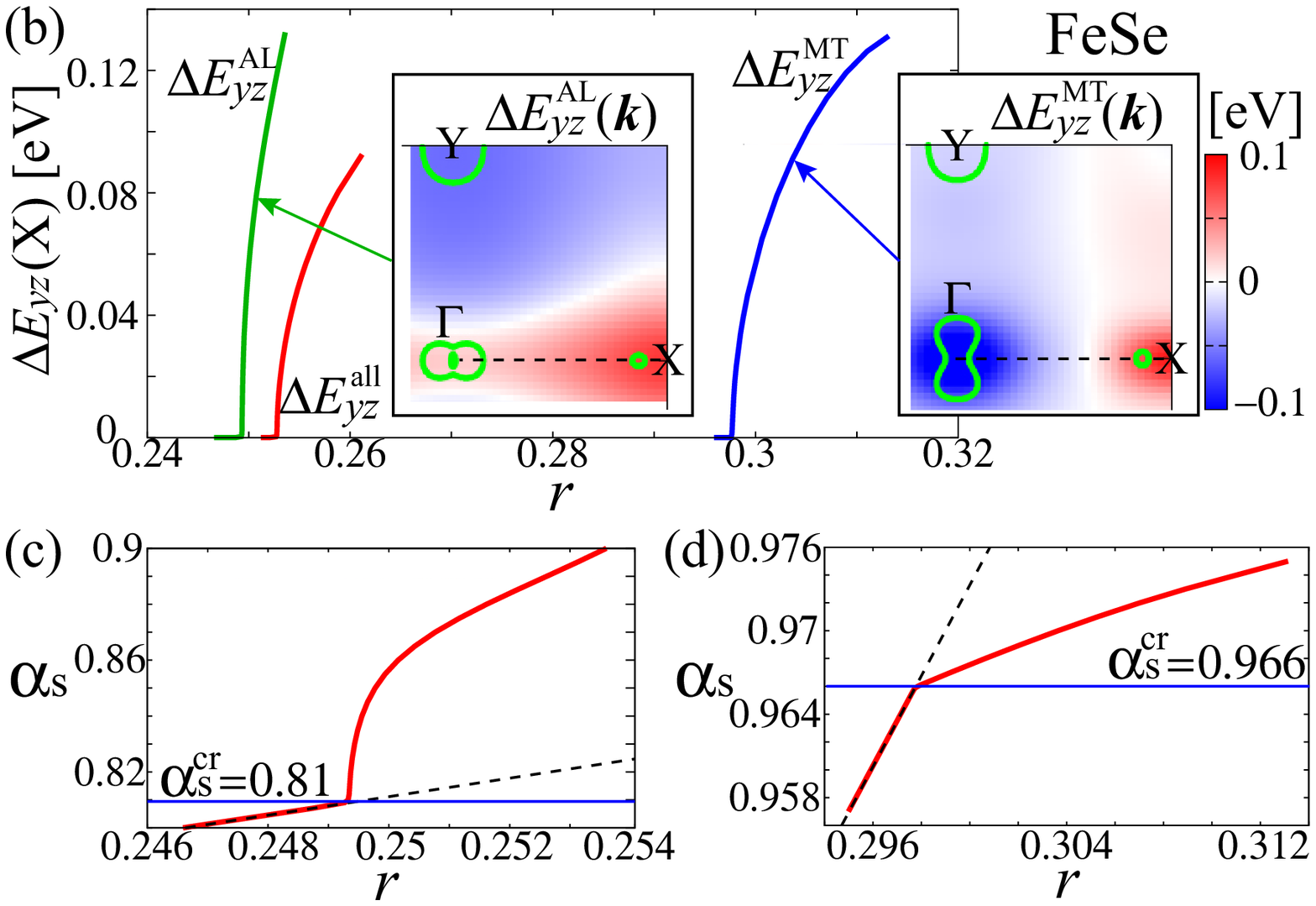}
\caption{
(color online)
(a) Expansion of the self-energy with respect to $\Delta E$.
The MT term (AL term) is included in ${\hat \Sigma}^{\rm MT(AL)}$.
(b) $r$-dependences of $\Delta E_{yz}^{\rm AL}$ and
$\Delta E_{yz}^{\rm MT}$ at X-point.
$\Delta E_{yz}^{\rm all}$ is the
orbital polarization derived from Eq. (\ref{eqn:Self}) 
shown in Fig. \ref{fig:FeSe} (c).
(c) $\a_S$ due to ${\hat \Sigma}^{\rm AL}$, and
(d) $\a_S$ due to ${\hat \Sigma}^{\rm MT}$.
}
\label{fig:ALMT}
\end{figure}

\begin{figure}[!htb]
\includegraphics[width=.8\linewidth]{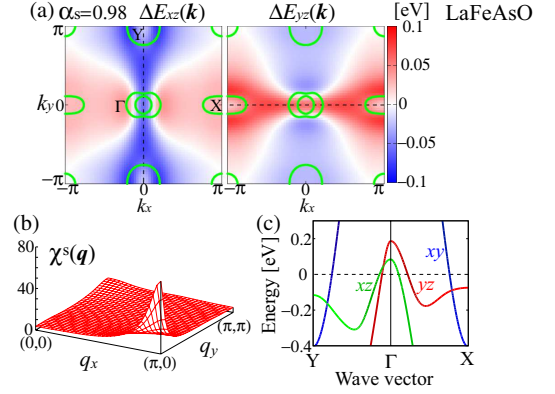}
\caption{
(color online)
(a) Obtained sign-preserving orbital polarization
($\Delta E_{xz}(\k), \Delta E_{yz}(\k)$) in the LaFeAsO model
for $\a_S=0.98$ ($r=0.376$).
The FSs are shown by the green lines.
The corresponding spin susceptibility and the bandstructure 
are shown in (b) and (c), respectively.
}
\label{fig:LaFeAsO}
\end{figure}

Here, we explain that the origin of the orbital order is 
the  positive feedback between
the orbital polarization ($\Delta E_{xz}(\k), \Delta E_{yz}(\k)$)
and the spin susceptibility.
For this purpose, we introduce the following two simplified self-energies
and perform the self-consistent calculations:
\begin{eqnarray}
&&\!\!\!\!\!\!
\Sigma_{l,l'}^{\rm AL}(k)=\Sigma_{l,l'}^{\rm H}
+T\sum_{q,m,m'} V_{l,m;l',m'}(q)G_{m,m'}^{\Delta\!E=0}(k-q),
\label{eqn:Self-AL}
\\
&&\!\!\!\!\!\!
\Sigma_{l,l'}^{\rm MT}(k)=\Sigma_{l,l'}^{\rm H}
+T\sum_{q,m,m'} V_{l,m;l',m'}^{\Delta\!E=0}(q)G_{m,m'}(k-q),
\label{eqn:Self-MT}
\end{eqnarray}
where the superscript ``$\Delta\!E=0$'' means the absence of the 
$C_2$ orbital polarization.
In Eq. (\ref{eqn:Self-AL}) (Eq. (\ref{eqn:Self-MT})),
only the feedback effect from the symmetry-breaking spin susceptibility 
(Green function) is included.
${\hat \Sigma}^{\rm AL(MT)}$ 
contains the Aslamazov-Larkin term (Maki-Thompson term) that is
the second-order (first-order) term with respect to $\chi^s$
in Fig. \ref{fig:ALMT} (a), which shows the 
expansion of the self-energy in Eq. (\ref{eqn:Self}) 
by using the relation 
${\hat G}={\hat G}^{\rm \Delta\!E=0}+ {\hat G}^{\rm \Delta\!E=0}
\cdot \Delta {\hat E}\cdot{\hat G}^{\rm \Delta\!E=0} + O(\Delta E^2)$;
see the SM C \cite{SM} for detailed explanation.
Figure \ref{fig:ALMT} (b)
shows the orbital polarizations $\Delta E_{yz}^{\rm AL}$ and
$\Delta E_{yz}^{\rm MT}$ at X-point derived
from Eqs. (\ref{eqn:Self-AL}) and (\ref{eqn:Self-MT}), respectively,
as functions of $r$ at $T=50$ meV.
Here, $\Delta E_{yz}^{\rm all}$ is the orbital polarization 
derived from Eq. (\ref{eqn:Self}), shown in Fig. \ref{fig:FeSe} (c).
The closeness of $\Delta E_{yz}^{\rm AL}$ and $\Delta E_{yz}^{\rm all}$
means that the orbital order is mainly driven by 
the Aslamazov-Larkin term ${\hat \Sigma}^{\rm AL}$.
Figure \ref{fig:ALMT} (c) shows that
$\a_S$ is strongly enlarged by the orbital order due to ${\hat \Sigma}^{\rm AL}$,
similarly to $\a_S$ in Fig. \ref{fig:FeSe} (e).
In contrast, $\a_S$ in Fig. \ref{fig:ALMT} (d)
is suppressed because of the wrong $\q=(\pi,0)$ nesting
on the $d_{yz}$-orbital due to ${\hat \Sigma}^{\rm MT}$.
Thus, the origin of the orbital order is 
the strong positive feedback between
the orbital polarization and $C_2$ spin susceptibility
described by the Aslamazov-Larkin term
in Eq. (\ref{eqn:Self-AL}).

Although ${\hat \Sigma}^{\rm AL}$ is the main driving force of the 
orbital order, the hole-pocket is elongated along the $k_x$-axis
in the ordered state driven by ${\hat \Sigma}^{\rm AL}$ 
as shown in the left inset of Fig. \ref{fig:ALMT} (b).
Therefore, the Maki-Thompson term 
is indispensable to reproduce the sign-reversing orbital polarization;
see the right inset of Fig. \ref{fig:ALMT} (b).
In SM C \cite{SM},
we explain analytically
why the Maki-Thompson term gives the sign reversal.
Thus, although $\Delta E^{\rm AL}_{yz}(k_x,0)$ is always positive
as shown in the left inset of Fig. \ref{fig:ALMT} (b),
$\Delta E_{yz}^{\rm all}(k_x,0)$ in Fig. \ref{fig:FeSe} (a)
given by Eq. (\ref{eqn:Self}) changes to negative at $k_x\sim 0$,
due to the sign-reversing Maki-Thompson term
that is small in magnitude at $\a_S\gtrsim \a_S^{\rm cr}$.

Next, we show the numerical results for the LaFeAsO model
with ${\hat z}^{-1}={\hat 1}$.
Then, the orbital order is realized for $\a_S>0.91$.
In Fig. \ref{fig:LaFeAsO} (a), 
we show the obtained orbital polarization at $\a_S=0.98$ and $T=47$ meV.
The e-FS1 around X-point is smaller than  
e-FS2 around Y-point due to the orbital polarization 
$\Delta E_{xz}({\rm Y})<0$ and $\Delta E_{yz}({\rm X})>0$.
In addition, $\Delta E_{yz}(k_x,0)$ ($\Delta E_{xz}(0,k_y)$)
is always positive (negative), and therefore
the outer hole-pocket is elongated 
along the $k_x$-axis.
This result is consistent with the Fermi surface deformation 
in the non-magnetic orthorhombic phase in NaFeAs
\cite{ARPES-NaFeAs}.
In the ordered state,
strong in-plane anisotropy of the spin susceptibility
$\chi^s(\pi,0)\gg \chi^s(0,\pi)$ is realized
as shown in Fig. \ref{fig:LaFeAsO} (b)
\cite{Kontani-softening},
consistently with the neutron scattering experiments.
The bandstructure is shown in Fig. \ref{fig:LaFeAsO} (c).

In this study, we succeeded in explaining 
(i) the small critical Stoner factor $\a_S^{\rm cr}$
and (ii) the relation $E_{xz}>E_{yz}$ near $\Gamma$-point in FeSe.
An origin of (i) is the smallness of the ratio ${\bar J}/{\bar U}$,
as verified by the SC-VC theory \cite{FeSe-Yamakawa}
and the renormalization group (RG) theory \cite{Tsuchiizu-Ru1,Tsuchiizu-Ru2}.
Another origin of (i) is the smallness of the FSs in FeSe,
since the three-point vertex 
${\hat \Lambda}_3\equiv \delta {\hat \chi}^0(\q)/\delta \Delta {\hat E}$ 
in the Aslamazov-Larkin term
increases when the particle-hole asymmetry is large 
\cite{FeSe-Yamakawa}.
The relation (ii) can be realized by the sign-reversing Maki-Thomson term.
Although the Maki-Thompson term is usually small, 
(ii) is actually realized when the FSs are smaller 
since the minimum of the Aslamazov-Larkin term 
$\Delta E^{\rm AL}_{yz}(\k)$ shifts to $\k={\bm 0}$.
Recently, the advantage of the small FSs for the nematicity
had been stressed by the RG study in Ref. \cite{Chubukov-RG}.


In summary, 
we investigated the $C_2$ symmetry breaking 
in the self-energy and susceptibility self-consistently,
and explained the experimental $C_2$-symmetric FSs and $\chi^s(\q)$.
In the FeSe model, experimental deformation of the FSs due to the 
sign-reversing orbital polarization is satisfactorily reproduced.
In the LaFeAsO model, in contrast,
spin fluctuations are strongly enlarged by the sign-preserving orbital polarization.
Thus, the key experiments below $T_{\rm str}$ are satisfactorily explained.
The orbital order originates from the positive feedback between the 
nematic orbital order and spin susceptibility due to the Aslamazov-Larkin term,
and the sign reversal of $\Delta E_{yz}(k_x,0)$ and $\Delta E_{xz}(0,k_y)$
in FeSe is caused by the Maki-Thompson term.

\acknowledgements
We are grateful to Y. Matsuda, T. Shibauchi, T. Shimojima,
A. Chubukov, J. Schmalian and R. Fernandes for useful discussions.
This study has been supported by Grants-in-Aid for Scientific 
Research from MEXT of Japan.




\clearpage

\makeatletter
\renewcommand{\thefigure}{S\arabic{figure}}
\renewcommand{\theequation}{S\arabic{equation}}
\makeatother
\setcounter{figure}{0}
\setcounter{equation}{0}
\setcounter{page}{1}
\setcounter{section}{1}

\begin{widetext}
\begin{center}
{\bf 
[Supplementary Material] \\
Sign-reversing orbital polarization in the nematic phase of FeSe \\
due to the symmetry-breaking in self-energy 
}%
\end{center}

\begin{center}
Seiichiro Onari$^{1,2}$, Youichi Yamakawa$^3$, and Hiroshi Kontani$^3$
\end{center}

\begin{center}
\textit{$^1$ Department of Physics, Okayama University, Okayama 700-8530, Japan}

\textit{$^2$ Research Institute for Interdisciplinary Science, Okayama University}

\textit{$^3$ Department of Physics, Nagoya University, Nagoya 464-8602, Japan}
\end{center}

\end{widetext}

\subsection{A: Eight-orbital models for FeSe and LaFeAsO,
effect of the spin-orbit interaction}

We introduce the eight-orbital $d$-$p$ models for FeSe and LaFeAsO
analyzed in the main text.
We first derived the first principles tight-binding models
using the WIEN2k and WANNIER90 codes.
Next, in order to obtain the experimentally observed Fermi surfaces (FSs),
we introduce the $\k$-dependent shifts for orbital $l$, $\delta E_l(\k)$,
by introducing the intra-orbital hopping parameters,
as we explain in Ref. \cite{FeSe-Yamakawa}.
We shift the $d_{xy}$-orbital band [$d_{xz/yz}$-orbital band] 
at ($\Gamma$, M, X) points
by ($0$, $-0.25$, $+0.24$) [($-0.24$, $0$, $+0.12$)] for FeSe,
and ($0$, $0$, $+0.05$) [($-0.05$, $0$, $+0.05$)] for LaFeAsO.
The unit is eV.

Next, we discuss the bandstructure 
in the presence of the spin-orbit interaction (SOI),
which is expressed as 
$\lambda \sum_i {\bm l}_i\cdot {\bm \sigma}_i$.
The matrix elements of ${\bm l}_i$ are given in Ref. \cite{Saito-SOI}.
To study the effect of the SOI, we have to use the ten-orbital model
since the ``unfolding'' is prohibited by the SOI.
In addition, the SU(2) symmetry for conduction electron spins
is violated by the SOI.
For these reason, in the presence of the SOI,
the numerical study of the self-energy calculation 
becomes very difficult.

\begin{figure}[!htb]
\includegraphics[width=.99\linewidth]{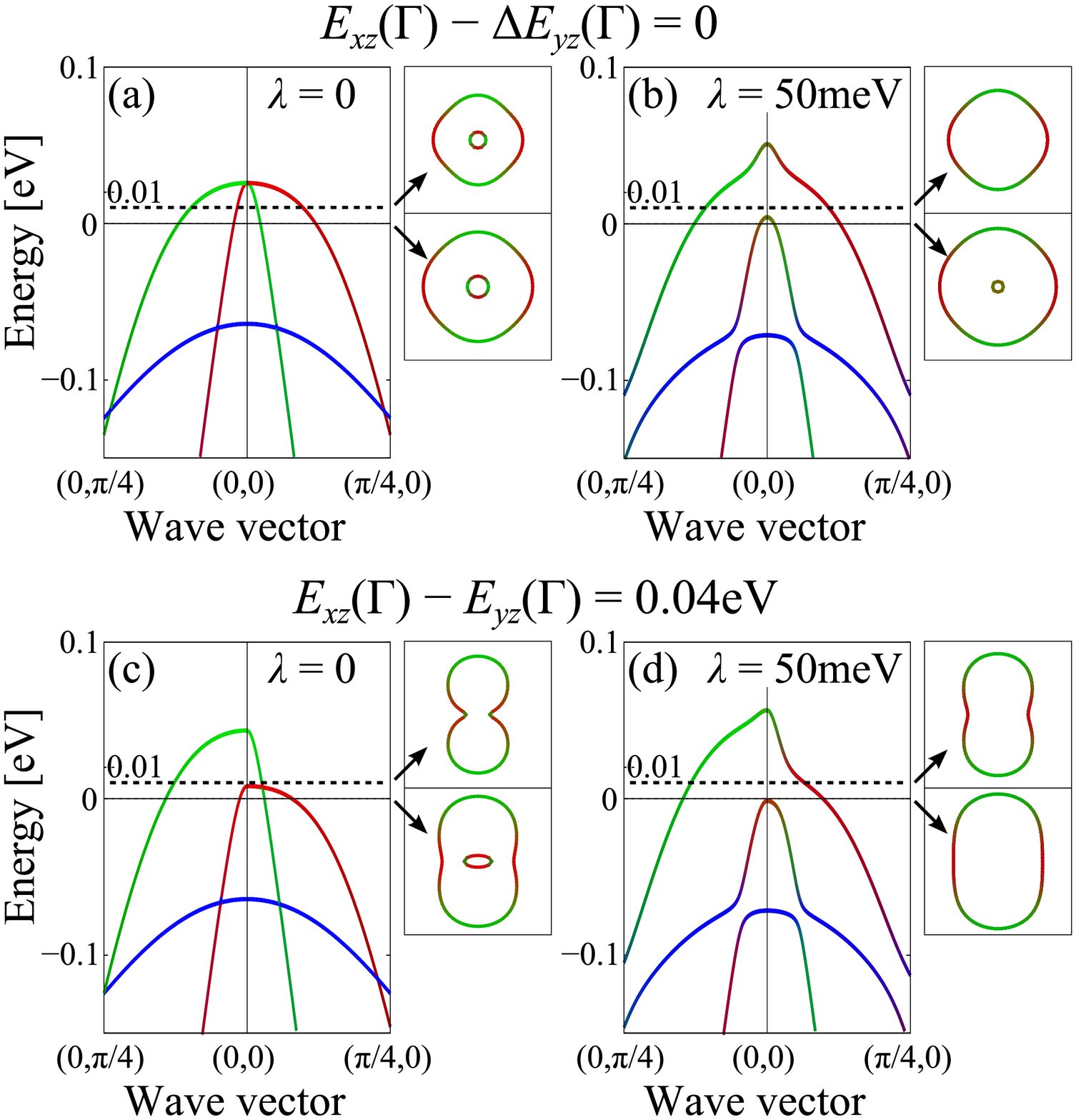}
\caption{
(color online)
Bandstructures and hole-FSs in the FeSe model.
(a) $\lambda=0$ (without SOI) and
(b) $\lambda=50$ meV (with SOI) for $E_{xz}-E_{yz}=0$ at $\Gamma$-point.
(c) $\lambda=0$ (without SOI) and
(d) $\lambda=50$ meV (with SOI) for $E_{xz}-E_{yz}=40$ meV at $\Gamma$-point.
}
\label{fig:bandSOI}
\end{figure}

Figures \ref{fig:bandSOI} (a) and (b) show the 
bandstructures and hole-FSs for $\lambda=0$ and $\lambda=50$ meV, respectively,
in the FeSe model in the absence of orbital polarization.
The main effect of the SOI on the hole-FSs is that the inner pocket
shrinks or sinks below the Fermi level.
In the presence of the orbital polarization  $E_{xz}-E_{yz}=40$ meV at 
$\Gamma$-point,
the bandstructures and hole-FSs are modified to 
Figs. \ref{fig:bandSOI} (c) for $\lambda=0$ and (d) for $\lambda=50$ meV.
In both cases, the outer hole-FS is mainly composed of the $d_{xz}$-orbital,
which is consistent with the experimental report \cite{FeSe-ARPES6},
when we set the radius of the hole-FS just slightly smaller 
than the present model ($\mu=+10$ meV).
The hollows of the hole-FS near the $k_x$-axis 
become smaller in the presence of the SOI.

Finally, the bare Coulomb interaction for the spin channel 
in the main text is
\begin{equation}
(\Gamma^{\mathrm{s}})_{l_{1}l_{2},l_{3}l_{4}} = \begin{cases}
U_{l_1,l_1}, & l_1=l_2=l_3=l_4 \\
U_{l_1,l_2}' , & l_1=l_3 \neq l_2=l_4 \\
J_{l_1,l_3}, & l_1=l_2 \neq l_3=l_4 \\
J_{l_1,l_2}, & l_1=l_4 \neq l_2=l_3 \\
0 , & \mathrm{otherwise}.
\end{cases}
\end{equation}
Also, the bare Coulomb interaction for the charge channel is
\begin{equation}
({\hat \Gamma}^{\mathrm{c}})_{l_{1}l_{2},l_{3}l_{4}} = \begin{cases}
-U_{l_1,l_1}, & l_1=l_2=l_3=l_4 \\
U_{l_1,l_2}'-2J_{1_1,l_2} , & l_1=l_3 \neq l_2=l_4 \\
-2U_{l_1,l_3}' + J_{l_1,l_3} , & l_1=l_2 \neq l_3=l_4 \\
-J_{1_1,l_2} , &l_1=l_4 \neq l_2=l_3 \\
0 . & \mathrm{otherwise}.
\end{cases}
\end{equation}
Here, $U_{l,l}$, $U_{l,l'}'$ and $J_{l,l'}$
are the first principles Coulomb interaction terms
given in Ref. \cite{Arita}.

\begin{figure}[!htb]
\includegraphics[width=.99\linewidth]{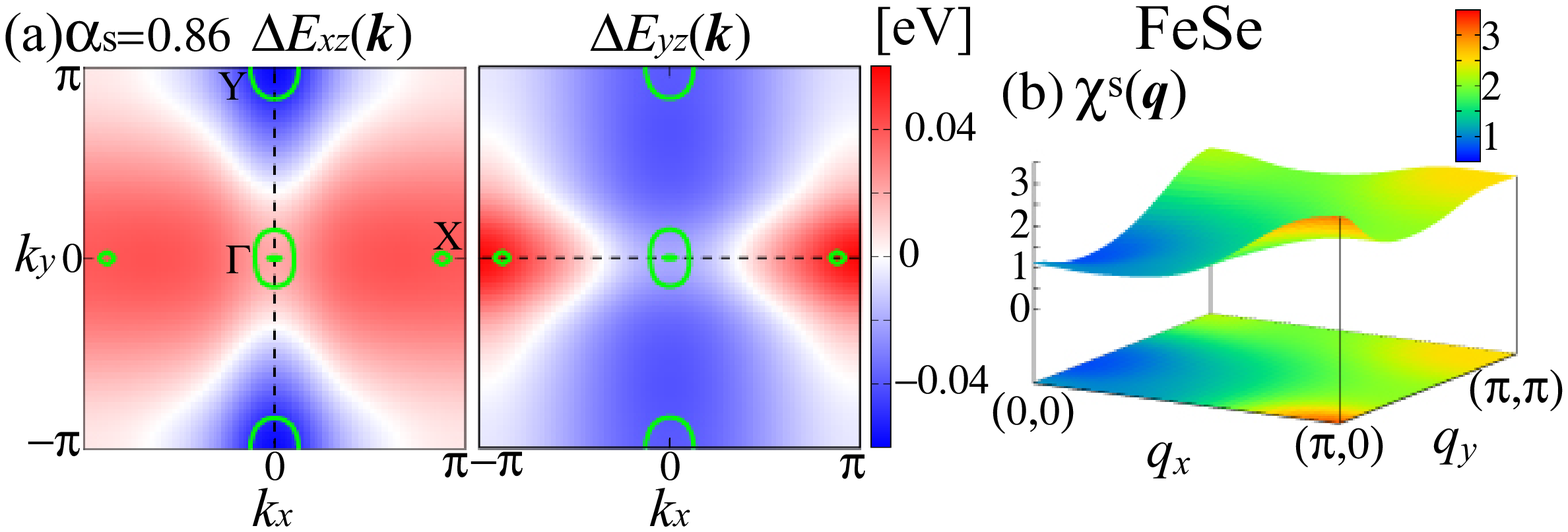}
\caption{
(color online)
(a) ($\Delta E_{xz}(\k)$, $\Delta E_{yz}(\k)$) and 
(b) $\chi^s(\q)$ in the FeSe model
for $\a_S=0.860$ ($r=0.506$) at $T=25$ meV in the case of $z=1/2$.
The obtained $C_2$-deformation of the FSs and that of $\chi^s(\q)$
are very similar to Figs. 2 (a) and (b) in the main text.
}
\label{fig:finite-z}
\end{figure}

In the main text, we put the renormalization factors as
$(z_1,z_2,z_3,z_4,z_5)=(1,1,1,1/1.6,1)$.
In the case of $(z_1,z_2,z_3,z_4,z_5)=z(1,1,1,1/1.6,1)$,
we proved that in Ref. \cite{FeSe-Yamakawa} that
essentially same numerical results are obtained by replacing
$r$ and $T$ with $r/z$ and $Tz$, respectively.
To verify this scaling relation,
we perform the numerical study for $z=1/2$:
Figure \ref{fig:finite-z} shows the obtained
(a) ($\Delta E_{xz}(\k)$, $\Delta E_{yz}(\k)$) and (b) $\chi^s(\q)$
in the FeSe model for $z=1/2$, $r=0.506$ and $T=25$ meV ($\a_S=0.860$).
These results are very similar to those for $z=1$, $r=0.257$ and $T=50$ meV
($\a_S=0.868$) shown in Fig. 2 in the main text.
Thus, the nematic orbital order is obtained by the present theory
even if the band renormalization effect is taken into account.

In the present study, we neglected
the damping of the quasiparticle 
given by the imaginary part of the self-energy.
We expect that this effect becomes small at sufficiently low temperatures,
around the realistic $T_{\rm str}\sim 100$ K.

\subsection{B: Orbital polarization in the single-hole-pocket FeSe model}

\begin{figure}[!htb]
\includegraphics[width=.6\linewidth]{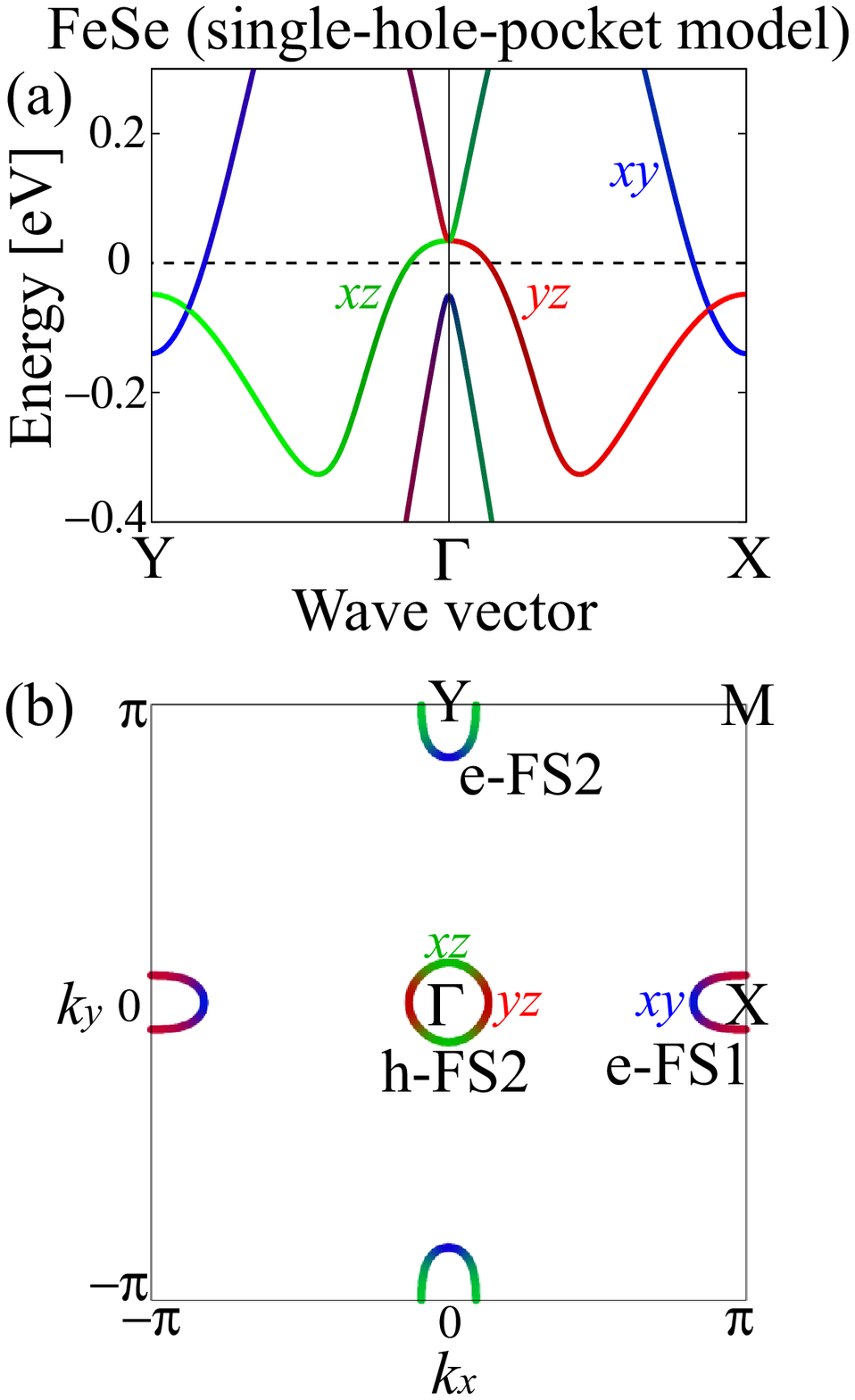}
\caption{
(color online)
(a) Bandstructure and (b) FSs in the single-hole-pocket FeSe model,
by shifting the $d_{xy}$-orbital level at $\Gamma$-point
by $-0.6$ eV.
}
\label{fig:sigle1}
\end{figure}

In FeSe, the main effect of the SOI on the hole FSs is that the 
inner pocket sinks under the Fermi level.
Similar bandstructure and the FS topology are realized 
by shifting the $d_{xy}$-orbital level at $\Gamma$-point by $-0.6$ eV,
as shown in Fig. \ref{fig:sigle1}.
Based on this ``single-hole-pocket FeSe model'',
we perform the numerical study of the orbital ordered state.

\begin{figure}[t]
\includegraphics[width=.99\linewidth]{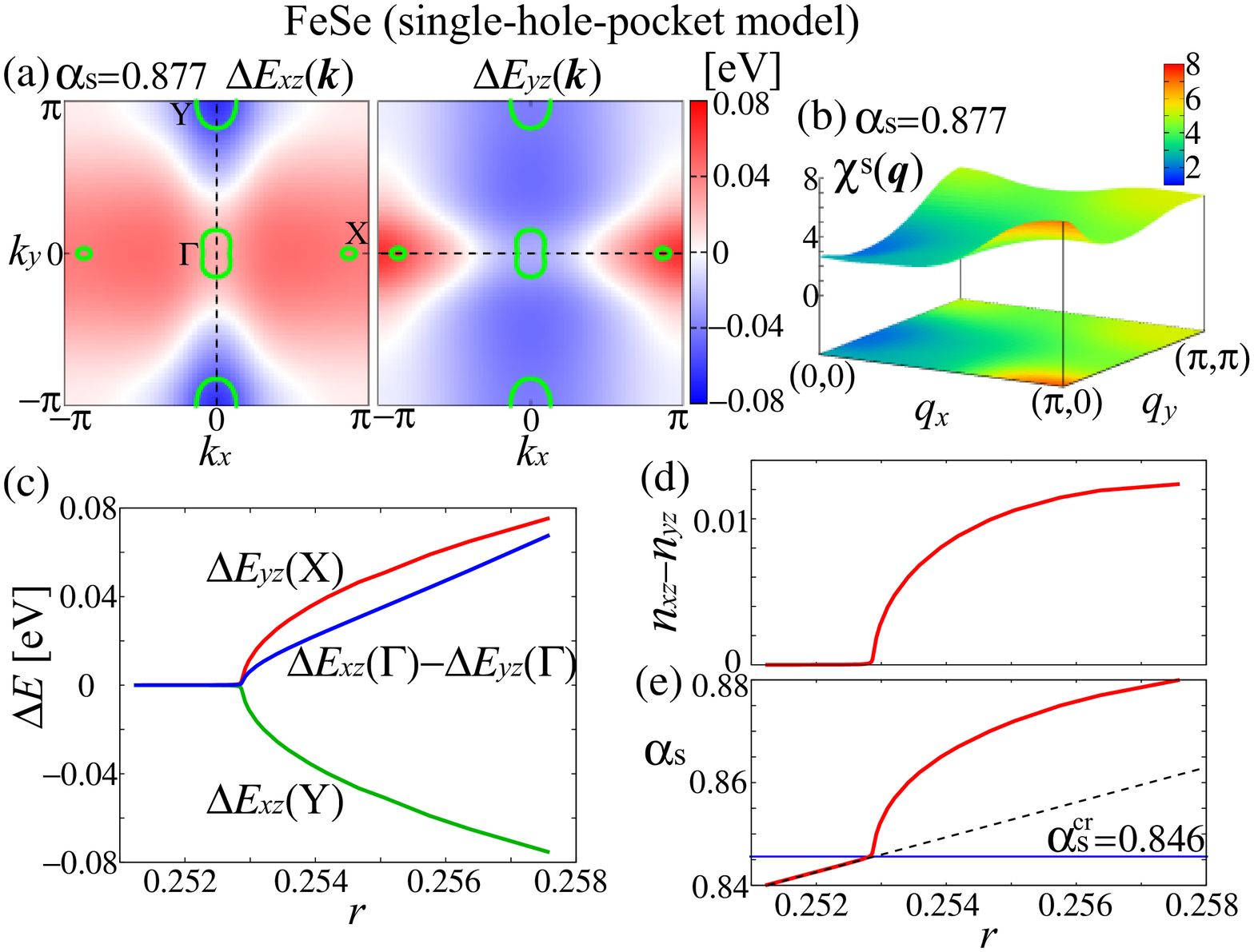}
\caption{
(color online)
(a) Obtained orbital polarization
($\Delta E_{xz}(\k)$, $\Delta E_{yz}(\k)$) and 
(b) spin susceptibility in the single-hole-pocket FeSe model 
at $\a_S=0.877$ ($r=0.256$).
The realized FSs are shown by the green lines.
(c) The orbital polarization,
(d) $\Delta n \equiv n_{xz}-n_{yz}$, and
(e) $\a_S$ for $0.258>r>0.251$.
}
\label{fig:single2}
\end{figure}

In the present model,
the nematic  orbital order is obtained for $\a_S>0.846$ ($r>0.253$).
In Fig. \ref{fig:single2}, we show (a) the orbital polarization
($\Delta E_{xz}(\k)$, $\Delta E_{yz}(\k)$) 
and (b) $C_2$ spin susceptibility $\chi^s(\pi,0) > \chi^s(0,\pi)$
obtained for $\a_S=0.877$.
Due to the orbital polarization,
the size of the e-FS1 around X-point is reduced, and 
two Dirac-cone Fermi pockets appear.
Another electron-pocket around Y-point, e-FS2, is enlarged.
In addition, the single hole-pocket is elongated along the $k_y$-axis.
Such FS deformation is essentially the same as that 
in the original FeSe model in the main text.
In Figs. \ref{fig:single2} (c)-(e),
we show the $r$-dependences of the orbital orders,
$\Delta n= n_{xz}-n_{yz}$, and $\a_S$, respectively.

\begin{figure}[!htb]
\includegraphics[width=.99\linewidth]{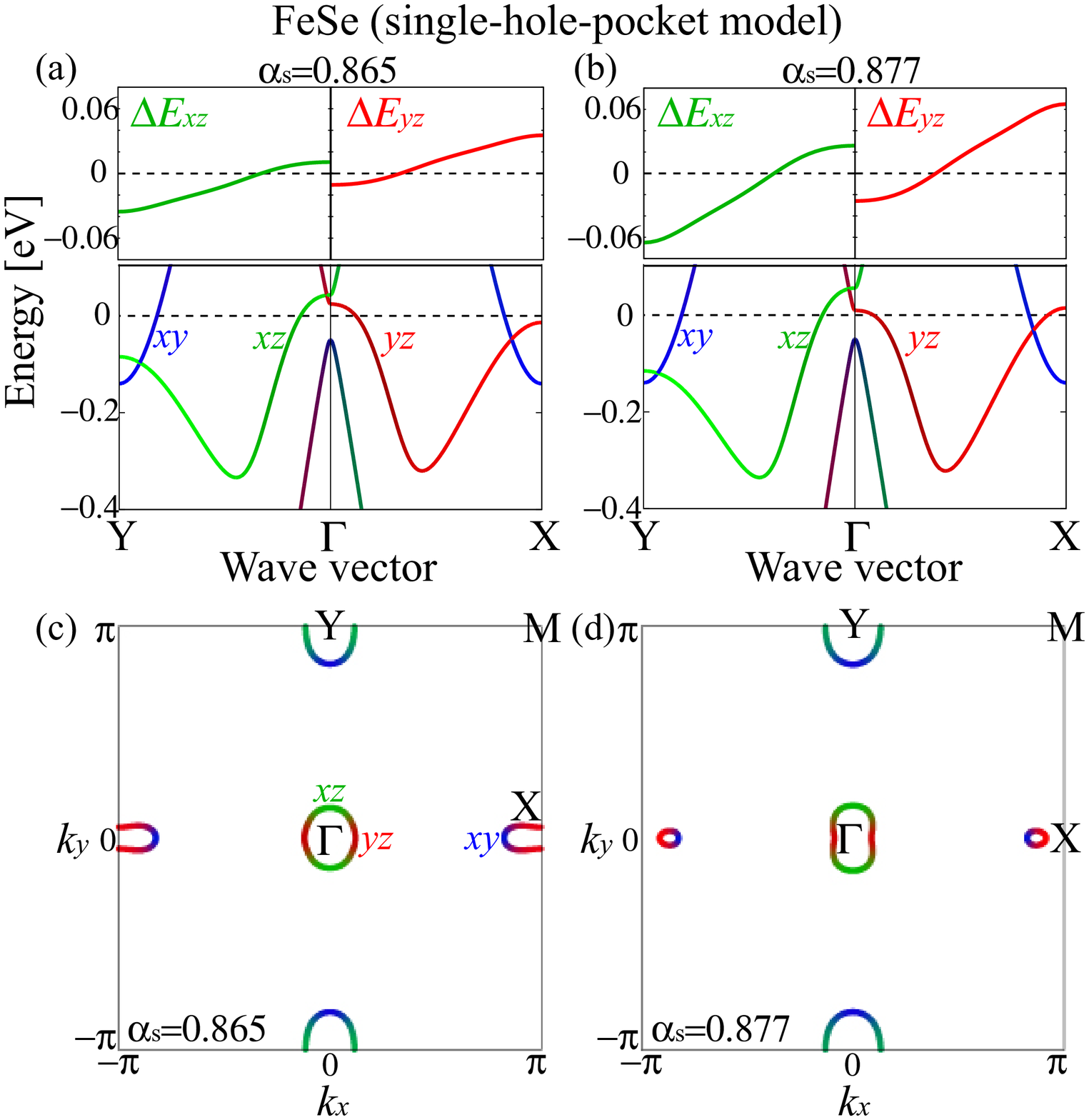}
\caption{
(color online)
Orbital polarizations and
bandstructures in the single-hole-pocket FeSe model along 
${\rm Y}\rightarrow \Gamma \rightarrow{\rm X}$
for (a) $\a_S=0.865$ and (b) $\a_S=0.877$.
The corresponding FSs are shown in (c) and (d), respectively.
}
\label{fig:single3}
\end{figure}

In Fig. \ref{fig:single3},
we display the obtained bandstructures and orbital polarization
for (a) $\a_S=0.865$ and (b) $\a_S=0.877$.
In both cases, 
$\Delta E_{xz}({\rm Y})-\Delta E_{yz}({\rm X})$ is negative, whereas
$\Delta E_{xz}(\Gamma)-\Delta E_{yz}(\Gamma)$ is positive.
The obtained FSs are also shown in Figs. \ref{fig:single3} (c) and (d), 
respectively.
These results are almost the same as the results
in Fig. 3 in the main text.

Thus, the sign-reversing orbital polarization in FeSe,
which is the main result of the main text, 
is also realized in the single-hole-pocket FeSe model.
This result indicates that the sign-reversing orbital order 
survives in the presence of the SOI.

\subsection{C: Comment on the self-consistent vertex correction (SC-VC) method}

In Ref. \cite{FeSe-Yamakawa},
the present authors studied various model Hamiltonians 
for Fe-based superconductors using the SC-VC theory.
In the FeSe model with $z_4^{-1}=3$,
The critical spin Stoner factor for the orbital order is
$\a_S^{\rm cr}=0.86$ at $T=50$ meV.
In the LaFeAsO model with $z_4^{-1}=1$ without the $d$-orbital level shifts,
$\a_S^{\rm cr}=0.975$.
These critical spin Stoner factors are slightly larger than those
obtained by the $C_2$ self-energy study in the main text.
In FeSe, the orbital order is realized 
even when the spin fluctuations are weak in FeSe,
due to the smallness of the ratio ${\bar J}/\bar{U}$ 
and the broadness of the $d_{xz/yz}$-orbital spin susceptibility,
as explained in  Ref. \cite{FeSe-Yamakawa}.

Here, we present an analytic explanation that 
the present self-consistent $C_2$ self-energy analysis is 
essentially equivalent to the SC-VC theory.
In the main text, Fig. 4 (a) shows the 
diagrammatic expression of the self-energy in Eq. (3),
by expanding the right-hand-side with respect to 
$\Delta E_l = \Delta \Sigma_l$: (i), (ii), and (iii), (iv)
corresponds to the Hartree term, the Maki-Thompson (MT) term, and 
the AL-terms, respectively.
In deriving Fig. 4 (a), we used the relation 
${\hat G}={\hat G}^{\rm \Delta\!E=0}+ {\hat G}^{\rm \Delta\!E=0}
\cdot \Delta {\hat E}\cdot{\hat G}^{\rm \Delta\!E=0} + O(\Delta E^2)$.
Based on Fig. 4 (a),
the linearized self-consistent equation for $\Delta E_l$ ($l=2,3$) 
is obtained by introducing the eigenvalue $\lambda_{\rm orb}$ as
\begin{eqnarray}
&&\lambda_{\rm orb}\Delta E_l(\k)
\approx T\sum_{k} \left[ U\{G_l(k)\}^2 \Delta E_l(\k) \right.
\nonumber \\
&& \ \ \ \ \ \ \ \left.+(2U'-J)\{G_{5-l}(k)\}^2 \Delta E_{5-l}(\k) \right]
\label{eqn:gap} \\
& &+ T\sum_{q}\frac32 V^s_l(q)\{G_l(k+q)\}^2 \Delta E_l(\k+\q)
\nonumber \\
& &+ T\sum_{q}(G_l(\k+q)+G_l(\k-q))\frac32 \{V^s_l(q)\}^2 
\Lambda_l(\Delta E; q), 
\nonumber
\end{eqnarray}
where the first two lines give the intra-orbital and inter-orbital 
Hartree terms, respectively,
and the third and fourth lines give the MT and AL terms, respectively.
Here, we assumed the orbital-diagonal ${\hat G}$ and ${\hat V}^s$ 
for $\Delta {\hat E}=0$, and
$\Lambda_l(\Delta E;q)= -T\sum_p \{G_l(p)\}^2G_l(p+q)\Delta E_l(\p)$
is the three-point vertex.
The eigenvalue $\lambda_{\rm orb}$ 
exceeds unity when the orbital order is realized.

First, we explain that the orbital order mainly 
originates from the AL term:
We put $\Delta E_{xz}(\k)=-\Delta E_{yz}(\k)=\Delta E$ 
to simplify the discussion,
and multiply both sides of Eq. (\ref{eqn:gap})
by $-T\sum_k \{G_l(k)\}^2$ by dropping the MT term.
Then, we obtain
\begin{eqnarray}
\lambda_{\rm orb} \chi^0({\bm 0})
\approx (2U'-J-U)\{\chi^0(\bm{0})\}^2 + X^{c}({\bm 0}) ,
\label{eqn:gap2} 
\end{eqnarray}
where $\chi^0({\bm 0})$ is the bare-bubble and
$X^{c}({\bm 0})$ is the charge-channel AL-VC for $l=2$ or $3$.
Thus, the orbital order ($\lambda_{\rm orb}=1$) is realized when
$X^c({\bm 0})> \chi^0({\bm 0})-(U-5J)\{\chi^0({\bm 0})\}^2 \sim 5J/U^2$
for $U=U'+2J$,
if we put $\chi^0({\bm 0})\sim 1/U$ for a qualitative estimation.
This result is consistent with the condition for $\a_C=1$ in the SC-VC theory
$X^c({\bm 0})= (U-5J)^{-1}-\chi^0({\bm 0}) \sim 5J/U^2$ for $J/U\ll1$
 \cite{FeSe-Yamakawa}.
Thus, the present theory
is a natural extension of the SC-VC theory
for the orbital-ordered state.

We also explain that the MT term gives the 
sign-reversing orbital polarization in FeSe 
as explained in the main text.
Considering the fact that $V_{3,3;3,3}^{\Delta\!E=0}(\q)$ has the peak at $(\pi,0)$,
the $\Delta E$-linear MT term in Eq. (\ref{eqn:gap}) for $d_{yz}$-orbital
at $\k\approx {\bm0}$ is approximately given as
%
\begin{eqnarray}
&&\!\!\!\!\!
-\sum_{\q} V_{3,3;3,3}^{\Delta\!E=0}(\q)\delta(E_{\k+\q}-\mu)\Delta E_{yz}(\k+\q)
\nonumber \\
&& \ \ \ \ \approx
-\left\{\sum_{\q} V_{3,3;3,3}^{\Delta\!E=0}(\q)\delta(E_{\q}-\mu)\right\}
\Delta E_{yz}(\pi,0),
\end{eqnarray}
where $E_{\k}$ is the $yz$-orbital band along $\Gamma$-X 
shown in Fig. 1 (a) in the main text, and we used the relation 
$T\sum_n (i\e_n +\mu-E_\k^\a)^{-2}= -\delta(E_\k^\a-\mu)$ at $T\approx0$.
Therefore, the MT term tends to induce the
sign reversal of $\Delta E_{yz(xz)}(\k)$ along the $k_{x(y)}$-axis,
as depicted in the right inset of Fig. 4 (b) in the main text.
Since $\Delta E^{\rm AL}_{yz}-\Delta E^{\rm AL}_{xz}$ 
is the smallest at $\Gamma$-point in FeSe,
the sign-reversing orbital polarization is realized by the MT term.

In the present $C_2$ self-energy study,
the higher-order MT- and AL-terms are included,
and therefore the obtained $\a_S^{\rm cr}$ is slightly smaller than 
that given by the SC-VC theory.
On the other hand, the feedback from the enhanced orbital susceptibility
given by the VC is absent in the $C_2$ self-energy study.
We will explain the relation between the $C_2$ self-energy theory 
and the SC-VC theory in more detail in future publications.

%
%
%
%
%


\begin{thebibliography}{99}
\bibitem{comment}
We verified that the ciritical value of $r$ for the orbital order
is slightly enlarged if Eq. (\ref{eqn:Self2}) is replaced with
$\Delta \Sigma_{l}(\k,\e_n)= 
\{\Sigma_{l,l}(k,\e_n)-\Sigma_{l}^{\rm A_{1g}}(\k,\e_n)\}$.
\end{thebibliography}
\end{document}